# Influence of Spattering on In-process Layer Surface Roughness during Laser Powder Bed Fusion

Haolin Zhang, Chaitanya Krishna Prasad Vallabh^, Xiayun Zhao*

ZIP-AM Lab, Department of Mechanical Engineering and Materials Science
University of Pittsburgh
Pittsburgh, PA, USA, 15213

^Current Address: Department of Mechanical Engineering, Stevens Institute of Technology,
Hoboken, NJ, USA 07030

*Author to whom the correspondence should be addressed
Email: xiayun.zhao@pitt.edu
Phone: (412)-648-4320

**Abstract**

Laser powder bed fusion (LPBF) based additive manufacturing (AM) holds great promise to efficiently produce high-performance metallic parts. However, LPBF processes tend to incur stochastic melt pool (MP) spattering, which would roughen workpiece in-process surface, thus weakening inter-layer bonding and causing issues like porosity, powder contamination, and recoater intervention. Understanding the consequential effect of MP spattering on layer surface is important for LPBF process control and part qualification. Yet it remains difficult due to the lack of process monitoring capability for concurrently tracking MP spatters and characterizing layer surfaces. In this work, using our lab-designed LPBF-specific fringe projection profilometry (FPP) along with an off-axis camera, we quantitatively evaluate the correlation between MP spattering and in-process layer surface roughness for the first time to reveal the potential influence of MP spatters on process anomaly and part defects. Specifically, a method of automatically and accurately extracting and registering MP spattering metrics is developed by machine learning of the in-situ off-axis camera imaging data. Each image is analyzed to obtain the MP's center location and the spatter count and ejection angle. These MP spatter signatures are registered for each monitored MP across each layer. Then, regression modeling is used to correlate each layer's registered MP spatter signature and its processing parameters with the layer's surface topography measured by the in-situ FPP. We find that the attained MP spatter feature profile can help predict the layer's surface roughness more accurately (> 50% less error), in contrast to the conventional approaches that would only use nominal process setting without any insight of real process dynamics. This is because the spatter information can reflect key process changes including the deviations in actual laser scan parameters and their effects. The results also corroborate the importance of spatter monitoring and the distinct influence of spattering on layer surface roughness. Our work paves a foundation to thoroughly elucidate and effectively control the role of MP spattering in defect formation during LPBF.

# 1. Introduction

Additive manufacturing (AM) attracts significant interests in recent decades due to its capability of fabricating parts with complex geometry. Laser powder bed fusion (LPBF) based AM is one of the most popular technologies in metal printing and adopted by various industries such as automobile, biomedical, and aerospace mainly for part prototyping [1-3]. Although widely used, LPBF based AM (LPBF-AM) faces challenges to be advanced toward industrial-scale additive production that demands process repeatability and part qualities. In LPBF manufacturing processes where laser sinters and fuses powder on the substrate or previous layers, there is complex physics interplaying among powder, laser, workpiece, and inert gas flow in chamber, inducing process anomalies like spattering and part defects like porosity and crack. Currently, researchers have developed various in-situ monitoring methods with a focus on capturing -melt pool (MP) behavior and morphology [4, 5]. However, most of these approaches are limited within a small region of interest (ROI) and a small number of layers. To establish comprehensive models of LPBF process-structure-property relationships for process and part qualification, more research is needed to monitor all types of phenomena in a continuous or near-continuous manner as well as quantify the effects of not only melt pools but also their associated spattering at a large scale across bigger area and through more layers.

## 1.1 Spattering phenomenon in LPBF-AM

Spattering is used to describe the material ejection from melt pools during LPBF printing process. A spatter can be categorized as a "droplet spatter" caused by vapor recoil pressure and Marangoni effect [6], or an un-sintered "powder spatter" caused by vapor-induced entrainment [7, 8]. A more detailed categorization of spatters is based on the formation mechanism, dividing into solid spatter, metallic jet, entrainment melting spatter (caused by gas flow), and defect induced spatter. In [9], the effect of scan speed on plume morphology and spatter generation is evaluated, showing that scan speed has direct impact on spatter formation. In [10], Schwerz et al. reported that the spatial location of the build and the gas flow direction in the build chamber are also key factors for spatter formation. The authors also studied the effect of layer thickness on the spatter formation, showing that larger (>80 μm) layer thicknesses had higher spatter count. Further, the authors inferred that the spatter re-deposition locations would primarily exhibit lack-of-fusion flaws. Researchers have also demonstrated the detrimental effects of spattering on LPBF printed parts since it could cause many defect modalities [11]. For instance, severe defects such as recoater streaking and lack of fusion flaws could be induced by high-rise spatters and oxidized spatters [10, 12-14]. All the existing research on various formation mechanisms and possible consequences of spattering in LPBF indicate a critical need for spatter monitoring and data analysis to understand and control its impacts on part properties during LPBF.

State-of-the-art monitoring methods, such as high-fidelity simulations and synchrotron X-ray imaging, have been used to identify spatter-induced defect formation mechanisms and features [15, 16]. However, these methods are expensive in terms of computation time and equipment cost. Therefore, researchers often would rather employ high-speed cameras to monitor the spatters and estimate their redeposited location [8, 10, 13, 17, 18]. But with such methods using relatively more affordable cameras (especially in contrast to high-end cameras and X-ray based equipment), it is difficult to trace the spatters completely from their ejection to landing during a LPBF printing process. Besides, these camera-based monitoring methods usually have limited ROI and cannot capture the spatter phenomenon fully in part scale. Overall, existing LPBF monitoring methods do not offer in-situ comprehensive spatter characterization capability due to the limited equipment accessibility, small field-of-view, or short monitoring duration.

On the other hand, research has emerged on analyzing available monitoring data for spatter features and their correlations to process characteristics and part properties. Repossini et al. [18] studied the effect of processing parameters on the spatter signatures for maraging steel prints in LPBF. They found that the spatter count is strongly related to the processing regime (i.e., conduction, transition, or keyhole). Also, they pointed out that the spatial location of spatters must be considered for evaluating their effects on part properties, especially for complex geometries. Similarly, Zhang et al. [17] found that spatter number, melt pool plume area, and plume orientation are directly correlated with the melt pool stability and the process parameters (laser power and scan speed). Besides, the authors found that the spatter orientation was not a direct indicator of process parameters. Moreover, spatters have been known to possibly interfere with the powder recoating for next layer, or be remelted during the print of subsequent layer(s), or persist across many layers [11]. Therefore, the behavior and evolution of landed spatters can play a significant role in determining the final forming of parts. In [19], authors studied the effect of spatter inclusions, which are larger than the powder layer thickness and could not be re-melted, using tensile test samples. These spatter inclusions occur more easily while printing un-sieved or re-used powder (> 5 cycles of reuse). It was also found that the spatter particle sizes were at least three times larger than the feedstock powders. The mechanical testing of these tensile samples revealed that the samples printed with re-used powders that tended to generate spatter inclusions had poorer tensile properties compared to the samples printed with fresh powders.

Overall, most of the existing research focus on characterizing the formation mechanism and dynamics of spatters. It remains unclear how spattering varies across each layer while printing a part and how the spatters would affect an in-process layer's surface roughness thus the printed part's properties, especially the internal defects (e.g., pores, cracks) that can be caused by the weak inter-layer bonding. This is mainly due to the limitations of current in-situ monitoring methods that cannot simultaneously provide spatiotemporally resolved spatter metrics and surface features. As such, there is a gap in developing capable multi-monitoring system and data analytics methods to acquire and analyze localized spatter signatures for detecting and predicting end-part properties.

*1.2 Surface roughness of LPBF prints: in-process layers and as-built parts*

To clarify, surface roughness of LPBF prints could refer to the surface roughness of workpiece in process, i.e., the in-process layer surface roughness due to the layer-by-layer processing nature of LPBF-AM, as well to the surface roughness of as-printed parts. Surface roughness of a printed part is known to be a critical part property related to the final mechanical performance such as fatigue life, because the exterior surface roughness can induce surface defects that initiate cracking and fracture. Researchers pay lots of attention and effort to characterize the as-built part's outside surface roughness. In [20-22], authors discovered the relationship between the processing parameters and the up-skin build surface roughness. All of these works reveal that with lower or higher energy density, the surface of the final part is rougher due to lack of fusion or material vaporization. Down-skin surface roughness is also studied in [23-25], and results reflect a weak correlation between processing parameters and down-skin surface roughness primarily due to decreased thermal conductivity of powder underneath melt pools. As the current surface roughness modeling or characterization work concentrates on the static exterior surface quality of a final printed part, there is a significant lack of studies on the dynamic interior layer surface roughness of a workpiece in a LPBF process. However, one should note that external roughness can be greatly reduced by post-processing such as machining and can no longer affect the end-use heavily, while the in-process layer surface roughness may induce internal defects that are hard to access for treatment or removal and can persist in the final parts. Particularly a rough in-process layer surface will weaken the inter-layer bonding and cause issues like porosity, powder contamination, and recoater intervention in LPBF. The in-process layer surface roughness during LPBF can cause severe, stubborn interior defects such as pores and cracks. Recently, researchers have used in-situ thermographic inspection to infer powder layer thickness based on a 1D thermal diffusion model and neural networks. Such an indirect thickness measurement method is prone to error and limited in resolution. Nevertheless, it has largely demonstrated the importance of knowing powder layer thickness for capturing possible causes and effects of rough powder surfaces such as catastrophic recoater crashes or abrasion and thermal distortions[26]. Therefore, elucidating the consequential effect of MP spattering on layer surface roughness is important for LPBF process control and part qualification, but has been overlooked and under-developed so far. It is desired to quantitatively understand the influence of spattering on layer surface roughness by correlating the in-situ monitored layer signatures of spatters and surfaces. Yet it remains difficult due to the lack of process monitoring capability for concurrently tracking MP spatters and characterizing layer surfaces with sufficient speed and accuracy.

### 1.3 Overview of this work

In this proof-of-concept work using a regular vision camera with relatively low sampling rate (compared to the laser scan speed) and limited resolution that can mainly observe large or high-temperature spatters which are melted, we aim to collect representative spatters and study the effect of powder agglomeration spatters and liquid spatters in LPBF processes. More tracked spatters and other types of spatters with smaller size, such as gas entrained spatter and powder spatter, can be studied in a similar framework as laid out in this work using a high-profile camera. Then, using our experiment data collected from a LPBF print of a multi-layer fatigue testing bars under different process settings, we conduct a correlation analysis between the spatter signatures and the layer's surface roughness metric. Further, to elucidate the significant effect of spattering on internal layer's surface roughness, we develop and compare different regression models that use an input of different combinations of spatter signature and processing parameters (laser power, laser scan speed, hatching) to predict the layer's surface roughness.

To accomplish this work, we first develop an off-axis camera-based monitoring system for imaging the spattering in situ during LPBF, and then establish a machine learning (ML) based framework for registering the monitored MP spatters. The methodological framework employs a segmentation neural network which segments each MP into background, MP center and spatter, and then register the essential spatter signatures including the MP's center location (coordinates), the counts of spatters, as well as each spatter's ejection angle relative to its MP center in the print layer using the physical printed part's coordinates system. The obtained data and results from our trained machine learning model are promising and agree with the existing studies performed on different materials and simpler geometries. Further, to address the issues mentioned in Sections 1.1-1.2, we investigate the uncharted effect of spattering on in-process layer surface property, using a cost-effective multi-sensor monitoring system including the off-axis camera and our previously developed fringe projection profilometry (FPP) [27]. Specifically, we analyze the influence of spattering on layer surface using regression models that correlate the MP spatter feature and the surface roughness calculated from the FPP measured surface topography.

The remaining of this paper is divided into the following sections. The experimental setup and design for spatter monitoring and signatures registration are elaborated in Section 2.1. Our in-situ two-sensor combined monitoring system - off-axis camera and FPP is employed as introduced in Section 2.2. Details about the machine learning model for spatter segmentation and feature extraction are included in Section 2.3. The methodology section ends with an explanation of the metric to be used for surface roughness characterization and the regression analysis to predict surface roughness from the processing parameters and in-situ monitored spatter signatures in Section 2.4. With a demonstration case of LPBF processing of 16 fatigue bars, we present the spatter registration and layer surface roughness measurement results in Sections 3.1 and 3.2. Sections 3.3 and 3.4 investigate the impact of laser scan parameters including power, scan speed, and hatching angle on spatter generation and layer surface quality. Section 3.5 evaluates the significance of spattering effect on layer surface roughness with a thorough correlation analysis using various combinations of nominal laser scan parameters and spatter signatures. Finally, Section 4 presents conclusive remarks on our work and recommendations for future improvement.

## 2. Methods

## 2.1 LPBF machine and experimental design

In this work, in-situ spatter monitoring and in-process layer surface measurement are performed during an experimental LPBF print of 16 standard fatigue test specimens (ASTM E466 standards) using Inconel 718 metal powder (particle size range - 20 to 60 μm) and a commercial DMLS machine - EOS M290. The schematic setup for this print is shown in Figure 1.The nominal layer thickness is 40 μm, and the hatching strategy adopts a 67 degrees rotation.

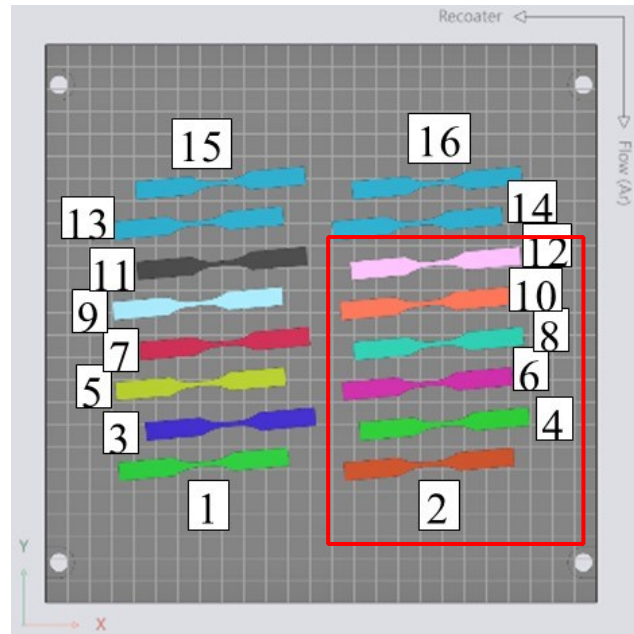

Figure 1. Schematic of our LPBF machine setup for printing 16 fatigue test bars. Bars enclosed in red box are used for surface roughness analysis through Fringe Projection Profilometry

One of the objectives of this experiment is to study the spatter phenomenon at different LPBF process regimes - conduction, transition, and keyhole, which are classified based on the research performed in processing parameters effect on MP geometry and morphology [5, 28]. The other objectives, which include the study of fatigue life, microstructure, and in-situ process signatures of the printed samples and their correlation, are reserved for other disseminations.

Corresponding to the three process regimes, the 16 fatigue bars are printed under various processing parameters, i.e., laser power (P), laser scan speed (V), hatching space (HS), as shown in Table 1. Also, two different scaling factors - Surface Energy Density (SED) and Volumetric Energy Density (VED) are calculated using Eq. (1) and Eq. (2) to account for the overall impact of laser scan parameters (P, V, HS) and nominal layer thickness ($t$). Researchers have been attempting to correlate SED and VED with end-part property such as up-skin surface roughness and porosity ratio [25, 29]. It is interesting to evaluate the performance of our method against existing methods. Therefore, in Section 3.5, we compare our layer surface roughness prediction models that use spattering signature with the common practice that would use VED. Contour scan is enabled for this experiment with laser power of 80 W and scan speed of 800 mm/s, and the internal printing time for one layer is around 120 second.

$$SED = \frac{P}{V \cdot HS} \quad (1)$$

$$VED = \frac{P}{V \cdot HS \cdot t} \quad (2)$$

Table 1. Our LPBF experimental process setting for printing the 16 fatigue samples

| Sample # | Power (W) | Velocity (m/s) | Hatching Space (μm) | SED (J/mm$^2$) | VED (J/mm$^3$) | Processing Regimes |
|---|---|---|---|---|---|---|
| 1 | 200 | 1.00 | 110 | 1.82 | 45.45 | Conduction |
| 2 | 250 | 1.00 | 110 | 2.27 | 56.82 | Conduction |
| 3 | 300 | 1.50 | 110 | 1.82 | 45.45 | Conduction |
| 4 | 250 | 0.75 | 110 | 3.03 | 75.75 | Transition |
| 5 | 285 | 0.96 | 110 | 2.70 | 67.47 | Transition |
| 6 | 300 | 1.00 | 110 | 2.73 | 68.19 | Transition |
| 7 | 350 | 1.00 | 110 | 3.18 | 79.55 | Keyhole |
| 8 | 200 | 0.50 | 110 | 3.64 | 90.91 | Keyhole |
| 9 | 250 | 0.50 | 110 | 4.55 | 113.6 | Keyhole |
| 10 | 300 | 0.50 | 110 | 5.45 | 136.4 | Keyhole |
| 11 | 200 | 1.00 | 80 | 2.50 | 62.50 | Conduction |
| 12 | 200 | 1.00 | 120 | 1.67 | 41.67 | Conduction |
| 13 | 250 | 0.50 | 80 | 6.25 | 156.3 | Keyhole |
| 14 | 250 | 0.50 | 120 | 4.17 | 104.2 | Keyhole |
| 15 | 200 | 1.50 | 110 | 1.21 | 30.30 | Conduction |
| 16 | 250 | 1.00 | 100 | 2.50 | 62.50 | Conduction |

*2.2 In-situ monitoring of LPBF process*

Figure 2 shows our lab-designed in-situ monitoring system, which is employed in this work to monitor spattering location and signatures as well as characterize in-process layer surface property.

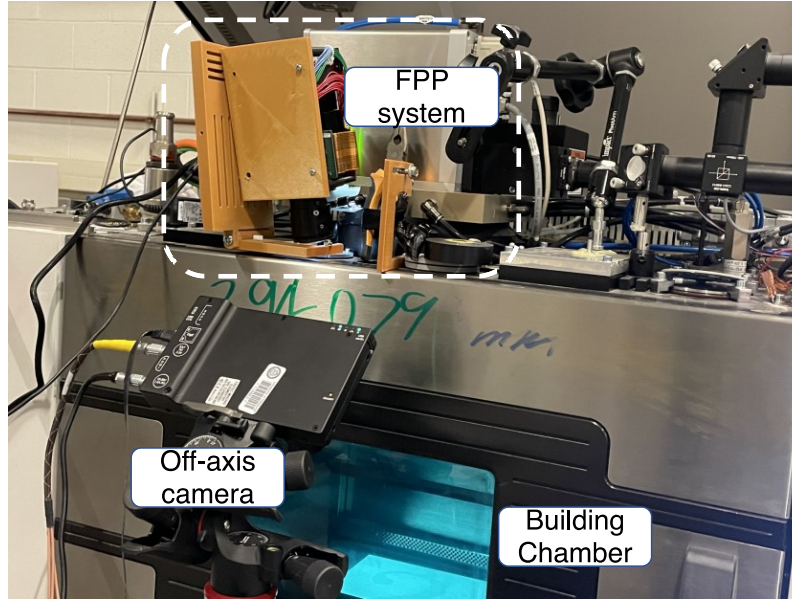

Figure 2. Our in-situ LPBF process monitoring system that integrates an off-axis camera and an in-house fringe projection profilometer for measuring the spatters and layer surface, respectively.

### *2.2.1 Off-axis camera-based spatter monitoring system*

The in-situ off-axis camera-based spattering monitoring is implemented to capture powder agglomeration spatters and liquid droplet spatters across the build plate. A high-speed camera (FASTEC IL5Q) is placed outside the build chamber of the LPBF printer (EOS M290), facing the building platform (Figure 2). This camera tracks the laser scan on each layer and captures images of laser and MP along with the build plate at a rate of 1,000 frames-per-second (fps). A representative image resulting from the off-axis camera is shown in Figure 3. The MP with large liquid droplet spatters is captured and used to extract information of MP coordinates on build plate as well as several spatter signatures including spatter ejection angle and counts. The acquired camera data is then analyzed using image processing methods. First, a perspective transformation is applied to correct the camera angle induced distortions. Details of the perspective transformation method are presented in our previous publication [30]. After perspective transformation, these images are segmented by a machine learning method as introduced in Section 2.3.

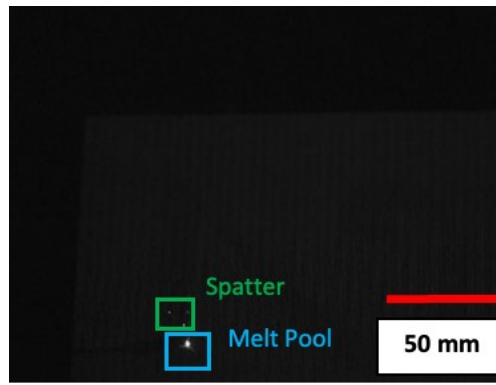

Figure 3. A representative spattering monitoring image captured by our off-axis camera with annotations of melt pool and spatter

## *2.2.2 Fringe projection profilometry (FPP) for in-process layer surface measurement*

FPP is an optical profilometry technology for characterizing the difference between the fringe pattern reflected from target object and the reflected fringe pattern from reference plane. It is widely used for measuring surface topography or part geometry in various fields. Preliminary work has shown that FPP can measure the in situ surface topography of a printed part on a layer basis [31, 32]. Recently, we have developed a LPBF-specific FPP system and method to improve the measurement accuracy by employing 2-dimensional Fast Fourier Transform (2DFFT) to assist the phase unwrapping process for unfolding the phase discontinuity [27]. Our in-house FPP setup (Figure 2) is comprised of the DLP optical projector (LightCrafter 4710 EVM G2, Texas Instruments, Dallas, TX) with a resolution of 1920 × 1080 pixels, a 12 Mega-Pixel CMOS camera (FL3-U3-120S3C-C, Flea3, Tele-dyne FLIR, Wilsonville, OR), and a computer unit to control and synchronize the system.

To quantify surface properties from the distorted fringe patterns, a FPP image pixel intensity value is encoded and transformed into phase value through a three-step phase shifting algorithm. The standard gamma correction is implemented to characterize the nonlinear sensor response between the camera and the LED projector. Wrapped phase of each pixel is calculated using Eq. (3), where $N$ is the total number of fringe patterns projected onto the target object.

$$\phi(x,y) = \arctan\left(\frac{-\sum_{i=1}^{N} I_i^{calibrated}(x,y)\sin(\delta_i)}{\sum_{i=1}^{N} I_i^{calibrated}(x,y)\cos(\delta_i)}\right) \tag{3}$$

The wrapped phase is then unwrapped using our new reference guided phase unwrapping method aided by 2DFFT [27]. A layer surface topography is derived by converting its unwrapped phase map to a height map using an experiment-calibrated ratio.

With the above-described FPP method, we can compute the in-process layer surface topography during LPBF and use it to further calculate the layer surface roughness *Sa*. Herein, the surface roughness is referred to as the arithmetic mean of the profile height deviation from the mean value. $Z(x,y)$ is the height deviation from the mean value at profile coordinates $x$ and $y$, as shown in Eq. (4).

$$S_a = \frac{1}{M \cdot N} \int_{y=1}^{N} \int_{x=1}^{M} |Z(x,y)| dxdy \tag{4}$$

## 2.3 A machine learning based framework of spatter monitoring data analysis for extracting and registering spatter features

### *2.3.1 Image segmentation via deep learning based segmentation neural network*

Segmentation neural networks have been used to extract and evaluate MP, plume and spatter information [33] but yield limited accuracy. In contrast, we adopt a state-of-art deep semantic segmentation neural network (NN), which is trained and tested based on DeepLabV3+ architecture with a variation of ResNet NN model [34] as an encoder backbone. Developed in [35], the DeepLabV3+ is a deep learning based neural network composed of encoder and decoder for the image segmentation purpose. Comparing to other convolutional NNs, Resnet prevents the gradient vanishing problem for training deeper NN by implementing residual connection to couple convolutional layers. Moreover, the NN structure utilizes a dilated convolution operation to effectively extract features from input image and upscale the output from encoder as the segmented output. The details of the dilated convolution are included in the Appendix (Section A1). The input to the NN is a spatter image acquired in-situ using the off-axis high-speed camera (Section 2.2.1) and the output is a segmented image with labels of background, spatters, and MP & plume. To exclude the noise from the large portion of background, the acquired off-axis camera images are first cropped to focus on the ROI. Specifically, our deep semantic segmentation NN is trained using 200 manually labelled images. To train the NN, the dataset is split into 80% for training and 20% for testing. The specific structure used for spatter segmentation is DeepLabV3+ with ResNet 18 as detailed in Table 2. The features are extracted through ResNet 18 convolutional operation. Then, the output from encoder is concatenated with the $1 \times 1$ convolution filtered input image and is further upscaled to the designed segmented output size.

Table 2. DeepLabV3+ with ResNet 18 encoder structure for training the spatter segmentation neural network

| Layer name | Output size | 18 layers |
|---|---|---|
| Conv group 1 | $112 \times 112$ | $7\times7$, 64, stride 2 |
| Conv group 2 | $56 \times 56$ | $3\times3$ max pool, stride 2 <br> $\begin{bmatrix} 3 \times 3, 64 \\ 3 \times 3, 64 \end{bmatrix} \times 3$ |
| Conv group 3 | $28 \times 28$ | $\begin{bmatrix} 3 \times 3, 128 \\ 3 \times 3, 128 \end{bmatrix} \times 2$ |
| Conv group 4 | $14 \times 14$ | $\begin{bmatrix} 3 \times 3, 256 \\ 3 \times 3, 256 \end{bmatrix} \times 2$ |
| Conv group 5 | $7 \times 7$ | $\begin{bmatrix} 3 \times 3, 512 \\ 3 \times 3, 512 \end{bmatrix} \times 2$ |

The performance of the trained NN is evaluated by a cross-entropy loss function as shown in Eq. (5), where $M$ is the number of possible classes of classified labels, $p_{oc}$ is the predicted probability that observation $o$ belongs to class $c$, and $y_{oc}$ is the actual probability that observation $o$ belongs to class $c$. After each epoch of training, the cross-entropy loss function is evaluated and used to optimize the model.

$$Cross\ Entropy\ Loss\ =\ -\sum_{c=1}^{M} y_{oc} log(p_{oc}) \qquad (5)$$

### *2.3.2 Spatter signatures extraction and registration*

To register the spatter signatures, each MP's center coordinates are determined after a perspective transformation using intensity-based thresholding method (elaborated in our previous publication [30]). With the ML based segmented images, we obtain spatter signatures associated with each MP, including the spatter count and spatter's ejection angle relative to laser scan direction. Then, the spatter signatures are registered by assigning them to the corresponding MP's center coordinates.

Specifically, after the machine learning based image segmentation, each pixel of in-situ monitored MP images is assigned with a specific label – (MP plume, spatter, background). Then, the Density-based spatial clustering of applications with noise (DBSCAN) algorithm is implemented to segregate pixels with spatter label into different clusters based on their density and spatial coordinates. The result from DBSCAN is the total count of spatters captured at this MP frame. The MP's center registration framework is further improved in this work by filtering the errors induced by the misclassified pixels. The areas of clustered groups that are formed by pixels with the label of MP core are compared; and the MP center coordinates are computed using the clustered group with the largest area. Using the segmented output from machine learning output, the spatter ejection angle is also characterized and registered as the spatter signatures. As shown in Figure 4, the spatter ejection angle is defined as the relative angle between the spatters and the laser scan direction which is set as reference.

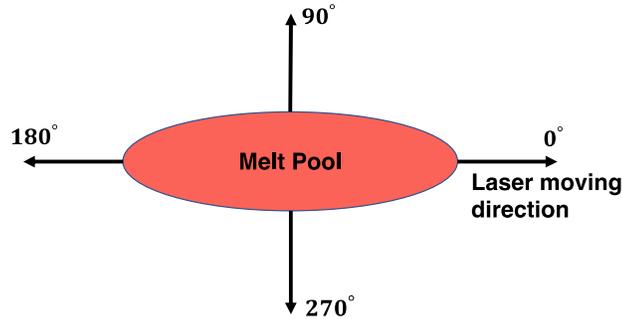

Figure 4. Illustration of the metric of spatter ejection angle

Overall, spatter quantity and spatter ejection angle are registered as two spatter signatures from the off-axis monitoring data using DeeplabV3+ image segmentation model and the DBSCAN clustering algorithm. The flow diagram (Figure 5) presents the spatter registration process of the work.

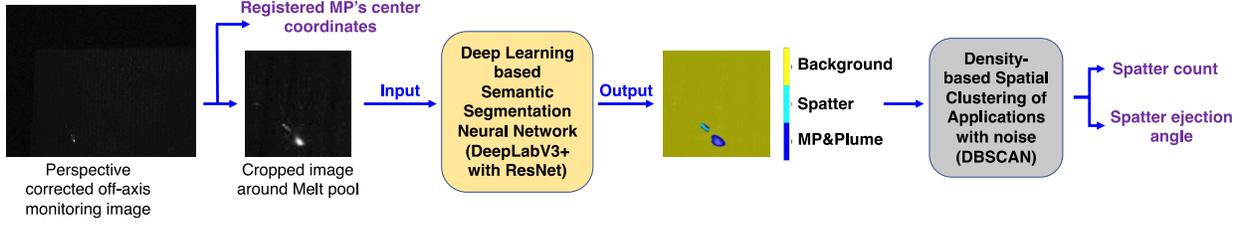

Figure 5. Flowchart of spatter features extraction and registration using off-axis camera-based in-situ monitoring data

## *2.4 Correlation analysis between spatter and in-process layer surface*

Support Vector Machine (SVM)-based regression models are employed to correlate the in-situ monitored spatter signatures and/or processing parameters with the in-process layer surface roughness. SVM is originally a classification method by defining margins [36]. First proposed in [37, 38], SVM shows proven accuracy in correlating sparse data with high dimension features. As a brief introduction, the SVM based regression model utilizes the sparseness from a SVM classification model and introduce an $\epsilon$ error function to replace the quadratic error function from a logistic regression model.

$$E_\epsilon(y(x) - \hat{y}) = \begin{cases} 0, & if\ |y(x) - \hat{y}| < \epsilon \\ |y(x) - \hat{y}| - \epsilon, & otherwise \end{cases} \quad (6)$$

As indicated in Eq. (6), the absolute error for a SVM regression model is calculated with the sensitivity factor $\epsilon$ between the regression model output $y(x)$ and target output $\hat{y}$. The sensitivity defines the error tube or boundary for the model, and any output at or out of the boundary are called as support vectors.

In this work, to quantitatively evaluate the significance of spattering among other potential factors - particularly the laser scan parameters, we develop different SVM models with various combinations of features and compare their performance. These SVM models use a Gaussian kernel filter to correlate the surface topography feature with in-situ monitored signature and/or specified process parameters. The input to the regression model is a subset of the following features: (1) laser power, (2) scan velocity, (3) hatching space, (4) laser scan angle derived from in-situ monitoring data, and (5) the average of in-situ registered spatter counts for each layer. The output is a corresponding layer's surface roughness. To obtain ground truth data for training and testing SVM models of the LPBF process and surface property relationships, we adopt the method as described in Section 2.2.2 to estimate surface roughness from the in-situ FPP measured layer topography. Our findings are reported in Section 3.5. The resulting insights on the influence of spattering will greatly facilitate the development of LPBF process monitoring and control technologies in the future.

## 3. Results and discussion

### *3.1 Registered spatter signatures*

As introduced in Section 2.3, spatter signatures are extracted and registered using the ML based image segmentation and spatial clustering methods. Specifically, a DeepLabV3+ plus ResNet-based segmentation NN is trained with a training dataset of 200 manually labeled images. The model is trained for 30 epochs to segment images into three different regions (background, spatter, and MP). The corresponding training accuracy of the model is shown in Figure 6, and a maximum accuracy of 99.57% is attained at the 147$^{th}$ iteration, which shows that the developed NN is accurate in segmenting images based on the provided labels. Representative segmented results using the developed ML aided method is shown in Figure 7.

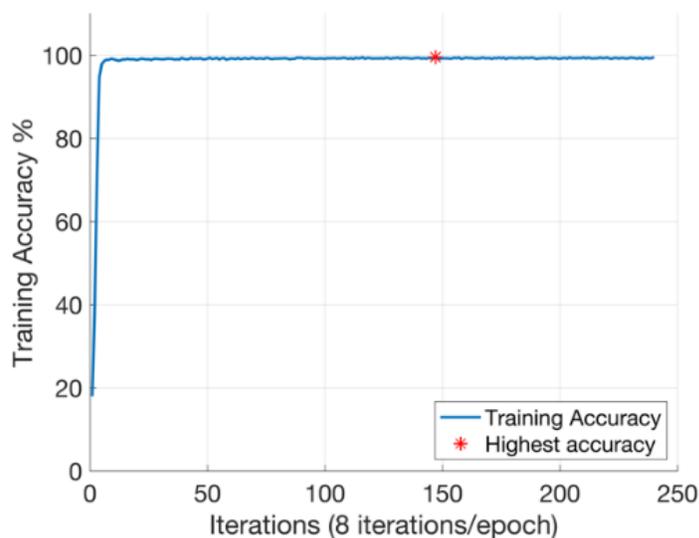

Figure 6. Training accuracy of the DeepLabV3+ plus ResNet-based spatter monitoring image segmentation method

The label of each pixel is color coded, as shown in Figure 7, where the spatter count is seen to be increasing with the VED value. Labeling errors could arise from possible human errors in the manual labeling process and thus cause misclassified pixels. However, the reason that manually labelled training data is implemented in this work is because the occurrence of noises presented in raw off-axis monitoring data, and it is difficult to differentiate the sparse and discrete spatters from other features (MP and background). As illustrated in Figure 8(a), the presented noises are a sequence of perpendicular spots as shown inside the marked red rectangle and caused by the camera sensor while acquiring data from outside the building chamber. These imaged vertical patterns could possibly be flare spot artifacts due to the high-luminance sources (laser and MP) causing intra-reflections within the camera elements that emerge at the film plane and form non-image information or flare on the captured image. In the future, such flare artifacts can be detected and removed from the spatter monitoring images using the setup reported in literature [39, 40]. With the appearance of lens flare artifacts, traditional image thresholding-based segmentation or clustering methods perform poorly in identifying spatters. Shown in Figure 8(b) and (c), the K-means clustering methods with different cluster numbers (K =3 and K=4) are tested on a representative MP image and fail to distinguish MP reflections from actual spatters. This failure is because the clustering methods are based on intensity level, which however does not change significantly among the spatter, noise, and plume. As compared in Figure 8, the segmentation result from the deep learning-based method of DeepLabV3+ plus Resnet (Section 2.3.1) along with our manually labeled training dataset successfully identifies the spatter by excluding the suspicious features from camera sensor noises.

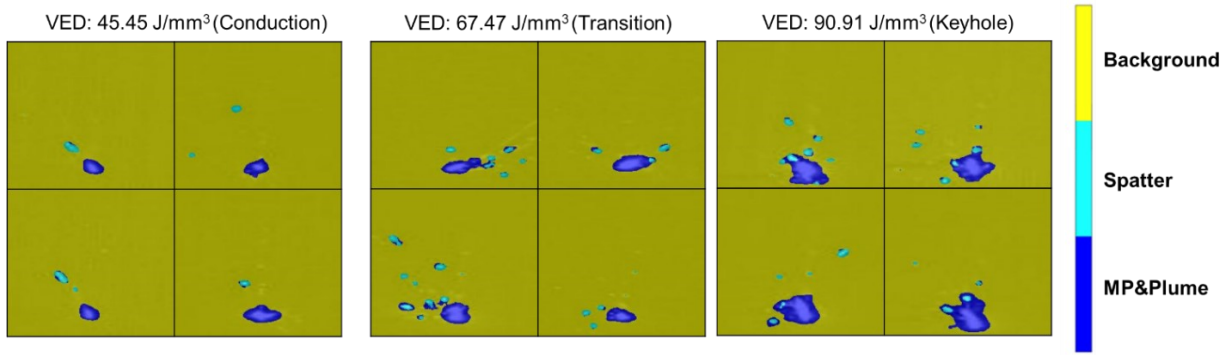

Figure 7. Representative segmentation results of four off-axis camera monitored images are shown for each of the three distinct LPBF process regimes (conduction, transition, keyhole).

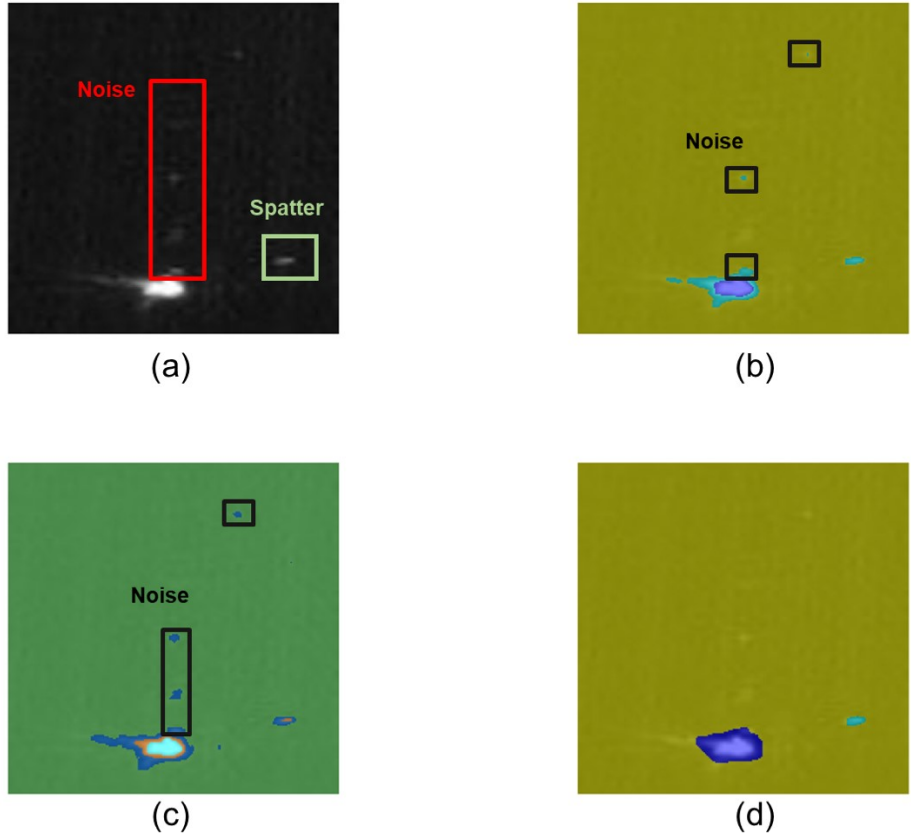

Figure 8. Compare the image segmentation performance of traditional clustering (K-means) and new deep learning-based methods. (a) Raw melt pool image captured by the off-axis camera with actual feature labels. (b) K-means image segmentation (K=3). (c) K-means image segmentation (K=4). Both (b) and (c) falsely classify the noises (e.g., flare artifacts) as spatters. (d) Deep learning (DeepLabV3+ with Resnet) -based Image segmentation used in this work extracts the correct feature of spatter.

After extracting the spatters through image processing, the corresponding spatter signatures such as spatter count, MP's center coordinates, and the ejection angle of spatters relative to laser scan direction are registered from the segmented images using the DBSCAN algorithm specified in Section 2.3. Figure 9 shows a representative result of registering the spatter count of each monitored MP at Layer 75 of all 16 Fatigue Bars (Figure 1). From the registered spatter signature map, the difference between contouring scan and hatching scan is reflected directly by the spatter count signature map with contouring scan exhibits low number of spatters (0~1). The registered spatter signatures maps build a comprehensive spatter dataset for subsequent quantitative analysis.

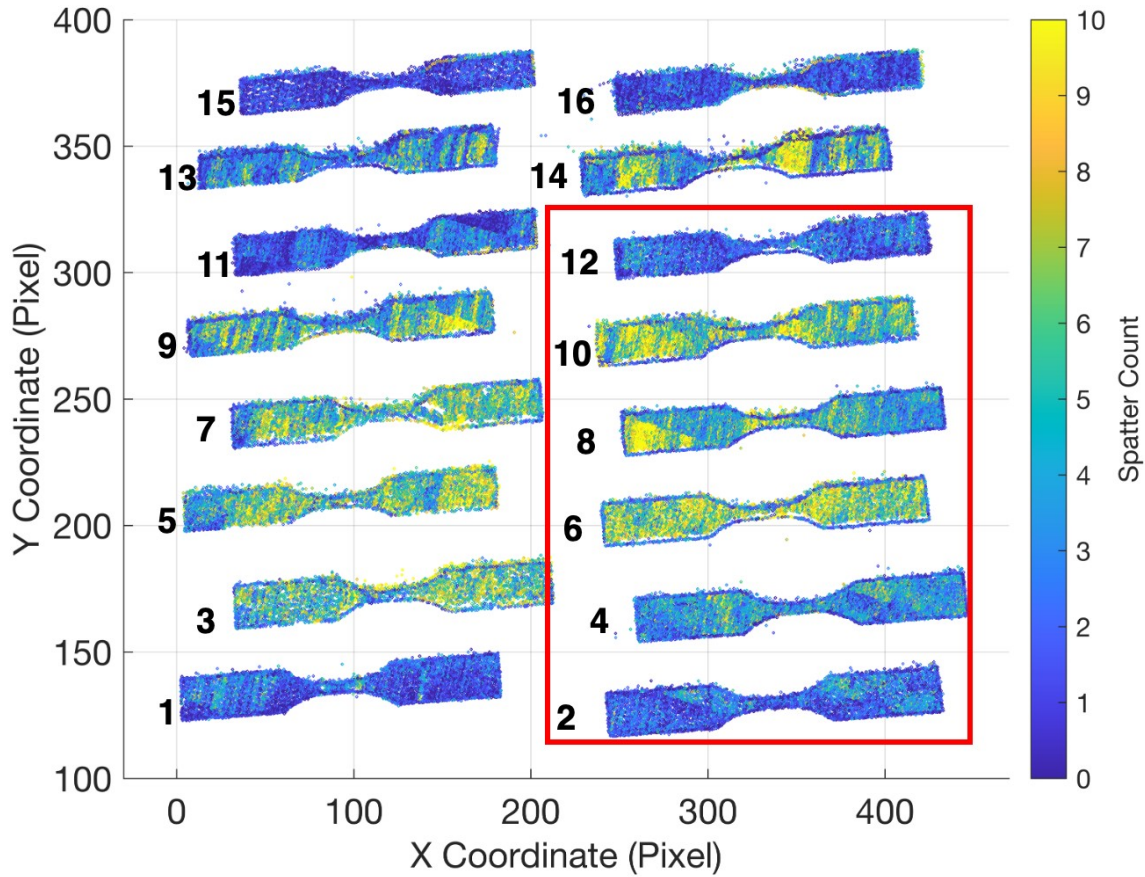

Figure 9. Representative result of spatter signature registration: the registered spatter count at each monitored melt pool in Layer 75 of all the 16 printed fatigue bars. The annotated numbers correspond to the processing parameters in Table 1. All the bars enclosed in the red box are monitored during the printing process by our in-situ Fringe Projection Profilometry and used for surface topography analysis and spatter-roughness correlation in subsequent sections.

### 3.2 In-process Layer surface topography and roughness

As introduced in Section 2.2.2, an in-situ 3-step phase shifting-based FPP method is applied to measure the in-process layer surface topography during LPBF processes. The surface topography is measured using the recoated powder bed surface as a zero-reference plane and calculated by analyzing the FPP images acquired right after the powder spreading and those acquired after the laser scan for that corresponding layer. Therefore, as shown in Figure 10, the measured height profile of a sample, in-process, layer displays negative height values, which are with respect to the unmelted powder bed surface and due to the laser fusion of powders and MP solidification shrinkage. Although our FPP system features a larger field of view than other literature reports, in this experiment with a limited hardware setup, the camera of FPP system captures only a portion of build plate, which only includes fatigue test bars 2, 4, 6, 8, 10, and 12 (as annotated in Figure 1 and Figure 9). Nevertheless, these monitored fatigue specimens fairly include two conduction regime samples (Fatigue Bars 2 & 12), two transition regime samples (Fatigue Bars 4 & 6), and two keyhole regime samples (Fatigue Bars 8 & 10). From the height profile (Figure 10), we see harmonic phase errors (strips in the background), which are induced by the limited number of steps (3 steps) used in the FPP method. To remove the harmonic errors, the FPP method will be improved in the future to incorporate efficient phase shifting algorithm with more steps or color encoded phase shifting.

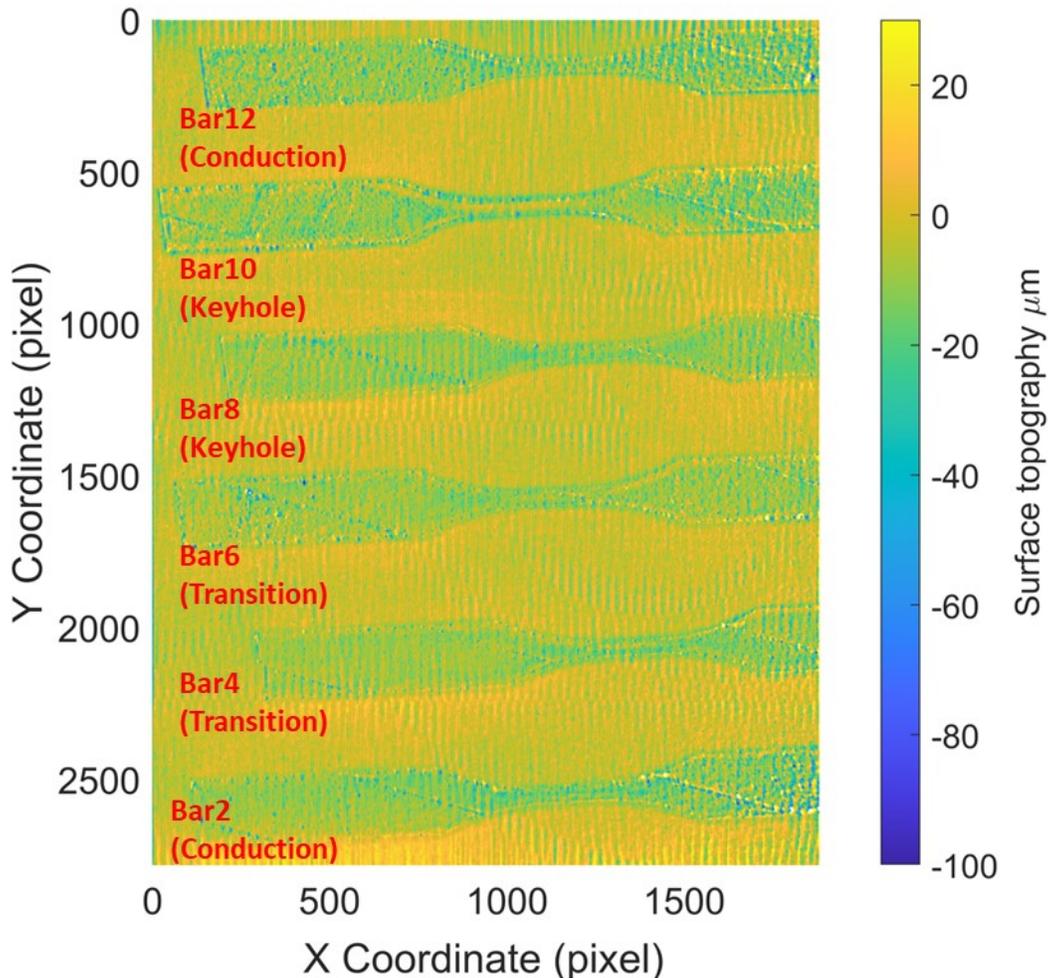

Figure 10. In-situ FPP measured surface topography of in-process Layer 66 of Fatigue Bars 2, 4, 6, 8, 10 and 12 as marked in Figure 1. Note the zero plane is the powder bed surface.

Further, to quantitatively compare the surface topography of different samples, an areal surface roughness is calculated using Eq. (4) in Section 2.2.2. Figure 11 shows a plot of our estimated surface roughness of Layer 66 of the monitored samples versus their processing VED. When the energy density is lower, lack of fusion occurs, and rough surface is induced due to un-melted or semi-melted powders on the surfaces. As the energy density increases, the surface roughness keeps decreasing as a result of complete melting. However, the surface roughness rises again as the energy density increases to the keyhole processing regimes where increased vapor/gas pressure could jet more droplet spatters out of the keyhole [41]. A similar pattern of how the surface roughness changes with VED through the three regimes is observed at an adjacent layer, i.e., Layer 67, as well, as shown in
Figure 12(a). Moreover, this "U-shaped" trend of our measured layer surface roughness against the VED can be explained by the literature finding [42] that spattering particle acceleration would linearly increase, decrease, and increase again with increasing line energy in the conduction, transition, and keyhole mode, respectively.

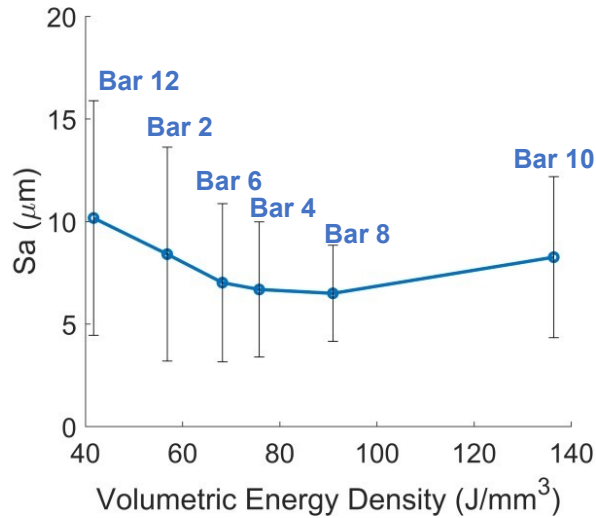

Figure 11. Surface roughness of in-process Layer 66 of the fatigue testing bars measured by the in-situ fringe projection profilometry against the corresponding processing regime's volumetric energy density. The error bar represents one standard deviation of $Z(x,y)$, which is the difference between the voxel height at profile coordinates $x$ and $y$ and the mean value as shown in Eq. (4).

As shown in Figure 12(a), our in-situ FPP measurement reveals that Layer 67 exhibits rougher surface than Layer 66 in each fatigue testing bar, although the two layers are adjacent within the same sample and printed under the same process setting. This phenomenon is consistent for all the monitored samples manufactured at various process regimes. The plausible reasons for this observed phenomenon will be reported in Section 3.4.

By further examining the registered spatter signatures acquired for the two layers (
Figure 12(b)), we find that the spatter quantities are closely related to the surface quality of the printed layer. As the numbers of spatter observed per MP increase, the in-process layer possesses more uneven features which induces irregular surface finish and high Sa value.
Our observations indicate a likely deficiency of the common practices that attempt to predict the surface quality of the printed part with process parameters or scaling factors such as VED. Spatter monitoring along with in-situ surface topography measurement can help capture the deviation or transition of process regimes and dynamic material behaviors more comprehensively during LPBF.

Figure 12 qualitatively reveals a complex relationship between the layer surface roughness, LPBF process setting (e.g., VED, hatching change across layers), and spatter count, necessitating a more in-depth quantitative analysis as presented in the following sections.

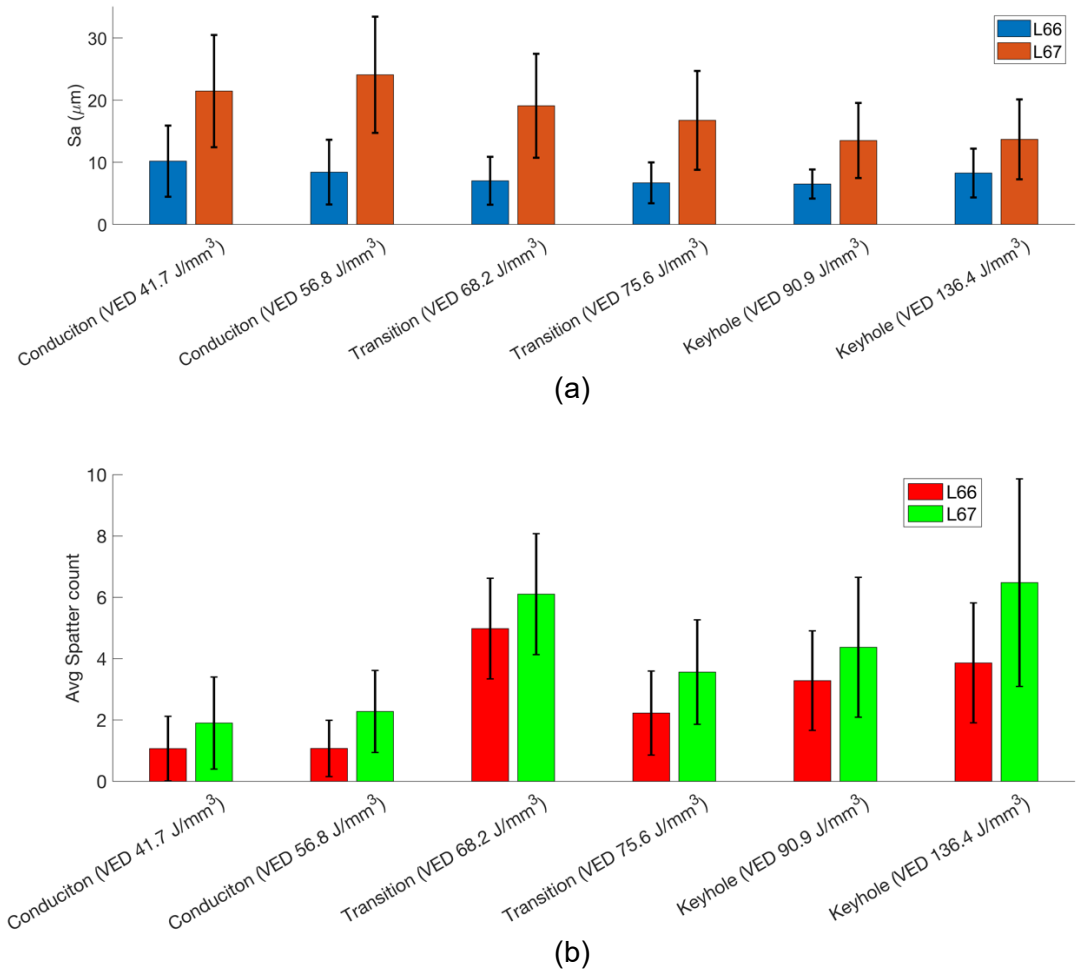

Figure 12. Surface roughness (a) and registered average spatter count per melt pool with one standard deviation (b) for Layer 66 and Layer 67. The error bar in (a) represents one standard deviation of $Z(x, y)$, which is the difference between the voxel height at profile coordinates $x$ and $y$ and the mean value as shown in Eq. (4).

### *3.3 Effect of laser power and scan speed on melt pool spattering*

In this section, we quantify the effect of laser power and scan speed on MP spattering and find a limited capability of using only laser power and speed for predicting spatter signatures. First, histograms of spatter count per MP and spatter ejection angle are plotted for a representative layer (i.e., Layer 66) at each of the three processing regimes, respectively (Figure 13). It can be observed that the spatter count distribution varies significantly with the processing regimes. As the VED increases, by average the conduction regime has the least number of spatters per MP (mostly 0-1), the transition regime has ~3 spatters per MP, and the keyhole regime has the most (~5) spatters per MP. Moreover, the spread of spatter count per MP in the keyhole regime is flatter (from 1 to 7) and wider than that in the other regimes, probably due to the stochastic pressure perturbation in the vicinity of MP. On the other hand, spatter ejection angle exhibits a consistent trend that most spatters are ejected from the MP tail (90º – 270º as denoted in Figure 4). This is because our off-axis camera mainly observes powder agglomeration spatters and liquid droplet spatters, which are caused primarily by the vapor jet from depression zone and the back surge of liquid MP. Since the spatter ejection angle distribution only differs slightly across the processing regimes, we decide not to consider it as an important feature in the subsequent process-spattering-surface correlation analysis, but only use the spatter count as a descriptive spatter feature.

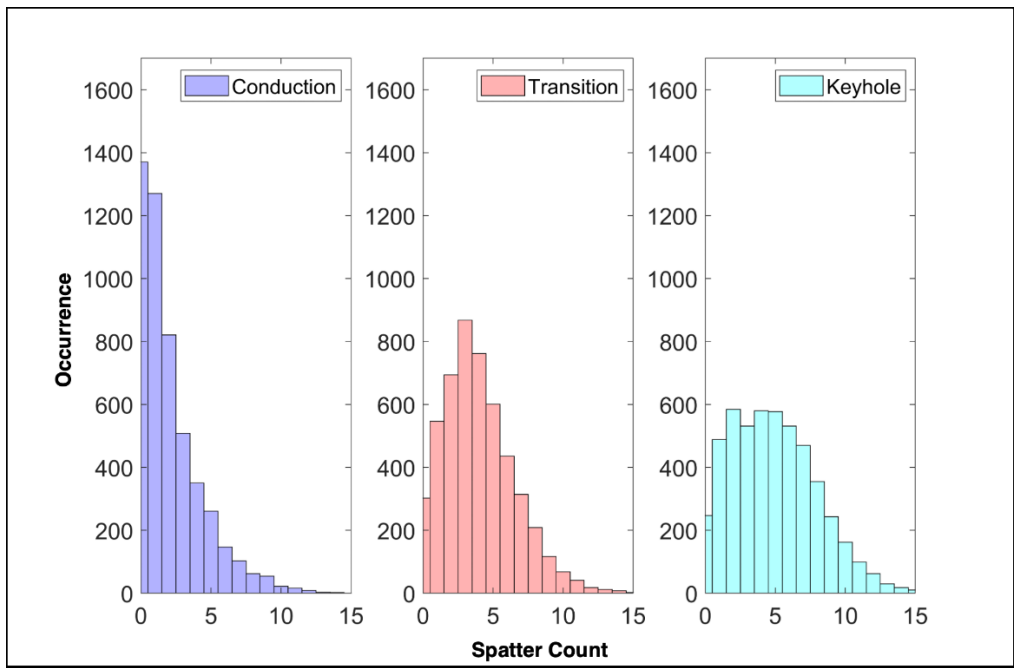

(a)

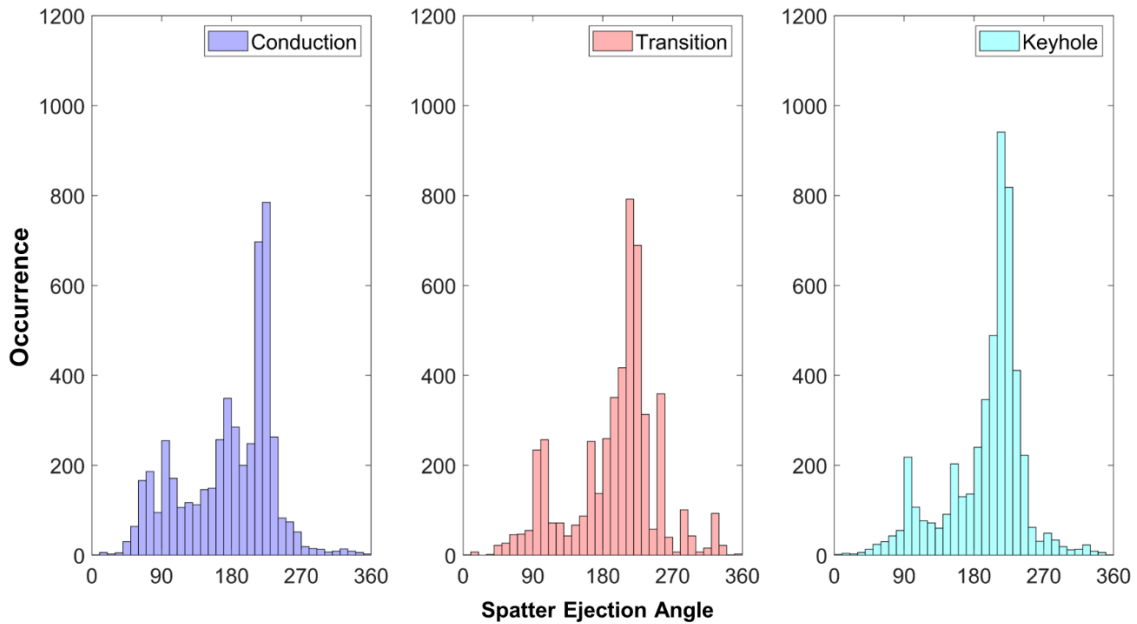

(b)

Figure 13. Histograms of registered spatter signatures for one representative layer monitored (L66) during a LPBF process in conduction regime (Fatigue Bar 2), transition regime (Fatigue Bar 4), and keyhole regime (Fatigue Bar 10). (a) Spatter Count; (b) Spatter ejection angle

By further using the registered spatter count data (Appendix Tables A-1 and A-2) for the sample layer – Layer 66 at different regimes, we construct regression models of the average spatter count in terms of laser power and scan speed, respectively, resulting in two linearly fitted curves as shown in Figure 14. Both regression models show a high $R^2$ value, indicating that the spatter quantity is highly correlated to the laser power and speed. The wide spread in Figure 13 of histograms and high error bars observed from Figure 14 do not necessarily indicate a lack of robustness in our spatter measurement. Because they are not uncommon especially in open-loop LPBF processes (as implemented in this work), which are not well controlled and subject to many possible variations including fluctuating laser absorption, instable melt pools, non-uniform material properties, and occurrence of porous defects. As such, the observations of spatter signature variations can rather be interpreted as a demonstration that our spatter monitoring can capture realistic process variations well.

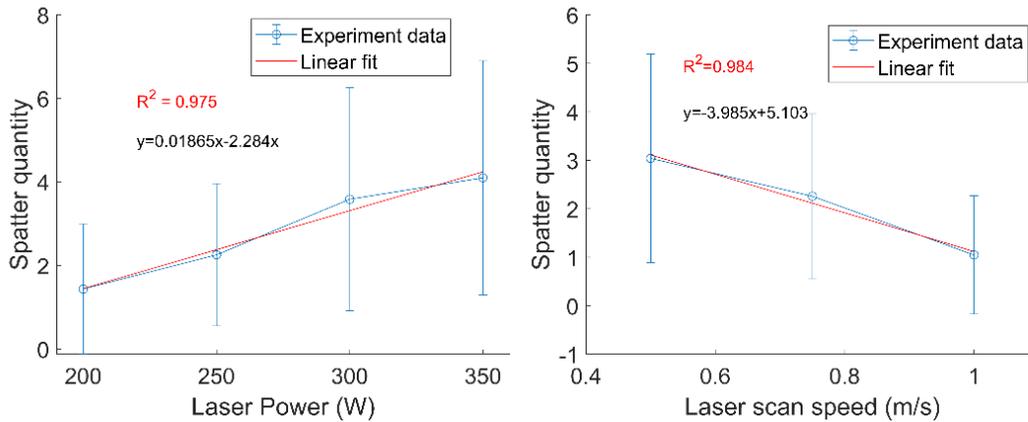

Figure 14. Linear curve fitting of the spatter count (quantity) with the laser power (left) and the laser scan speed (right). The blue circles are the mean values for Layer 66 of the samples printed in the specified parameter and the error bars indicate the standard deviation

The relationship between spatter count and processing parameters is shown to be linear but only observed within the same layer. When we examine the average spatter count per MP at different layers of three sample fatigue bars (Fatigue Bars 2, 4, and 8) as shown in Figure 15, it is found that both the spatter count and its relationship to the power and speed (i.e., the linear model coefficients) would change across the layers. This observation indicates that the spatter formation is not only attributed to processing parameters but also subjected to other factors such as contour scan, hatching space, layer thickness, and build locations. Specifically, the outlier layers that have obviously more spatters per MP are found from Figure 15 to be Layer 67 and Layer 75, which are suspected to be associated with a distinct hatching angle (elaborated in Section 3.4), since all the other process conditions (power, speed, hatching space, layer thickness, etc.) nominally remain the same as the other layers. Therefore, using solely nominal laser power and speed cannot fully reflect or accurately predict the spattering phenomenon, especially in a practical scenario of printing multi-layer parts. This necessitates the in-situ monitoring of spatters whose features are shown in this work to be capable of capturing real process variations in not only laser power and speed, but also other potential factors as presented in the next section.

Meanwhile, it is worth noting that our measured spatter quantity displays consistent curve fitting models with similar slopes and intercepts (i.e., similar spatter count values) for the layers, e.g., Layer 69 and Layer 72 as shown in Figure 15, which have nearly identical process conditions including a similar hatching angle of 95° and 116°, respectively, as detailed in next section. This good agreement between spatter quantity measurement and identical (nominally) process conditions demonstrates the reproducibility of our spatter monitoring and measurement methods.

Moreover, we find that the spatter occurrence has a wide range of distribution under different processing regimes. This emphasizes the process variation due to the hatching pattern. Also, it should be noted that spatter is a complex product associated with MP geometry, plume shape, location of the build, and gas flow condition, which tend to vary among laser scan vectors. The effect of such variation as observed on localized property will be investigated in our future work. In this work, a general part-level average signatures are used, and layer-wise difference observed during the printing is evaluated and studied.

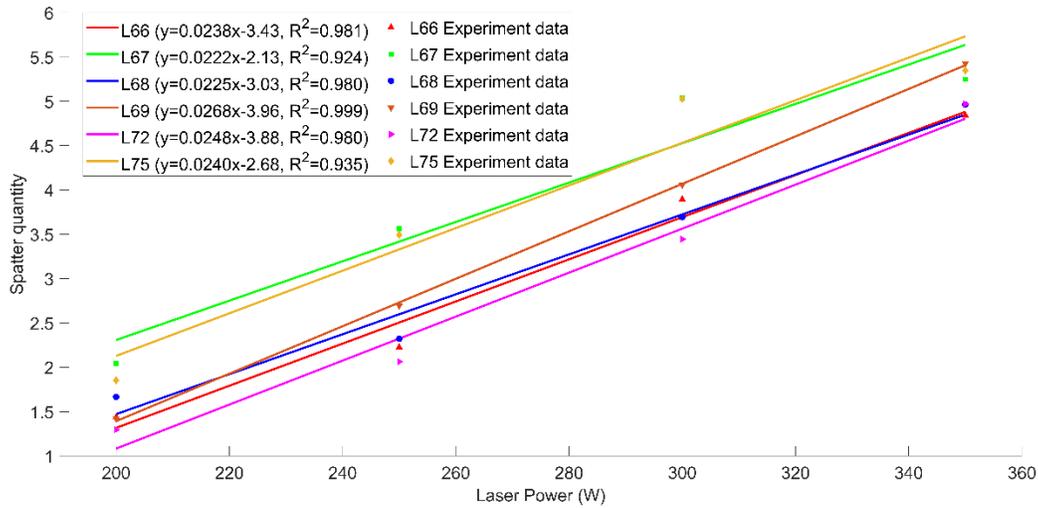

(a)

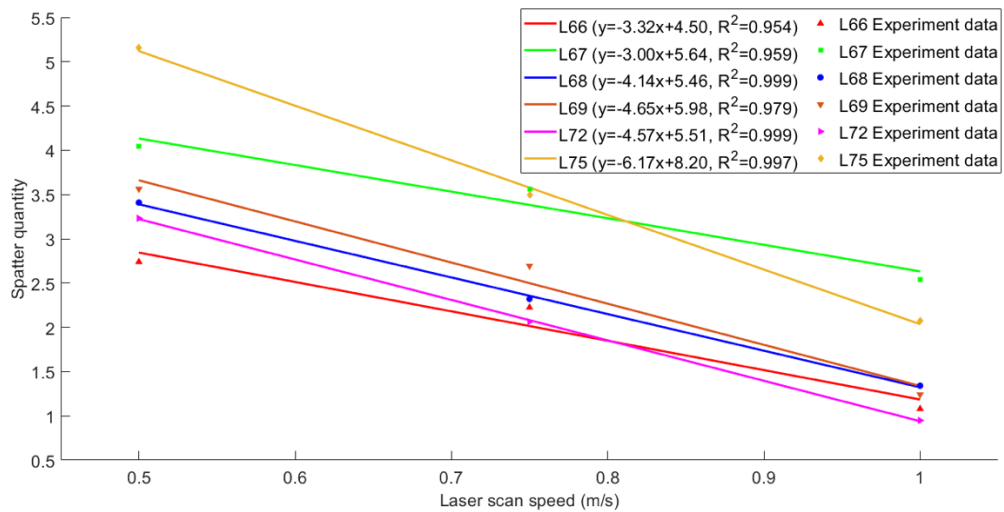

(b)

Figure 15. Average spatter count per melt pool (MP) and its relationship with laser power (a) and speed (b) can vary significantly across layers due to potential changes in the process conditions like hatching angle even though the nominal settings of laser power and speed remain the same. This implies in-situ monitored spatter features could reveal process variations in not only laser power and speed but also other factors (e.g., hatching rotation). The datapoints are calculated as the average spatter count per MP using our registered spatter count values of all monitored MPs at Layers 66-75. Fatigue Bars 1, 2, 6, and 7 are used for correlation between spatter and laser power (Bars manufactured with constant laser speed but varying laser power).

Fatigue Bars 2, 4, and 14 are used for correlation between spatter and laser scan speed (Bars manufactured with constant laser power but varying scan speed)

### *3.4 Impact of hatching angle on melt pool spattering*

In this section, the influence of hatching angle on spatter generation and surface roughness is studied. Herein, the hatching angle at each layer is defined as the angle between the laser scan vector and the horizontal axis. As detailed in Section 2.1, a hatching pattern with a rotation angle of 67° is implemented for the 16-fatigue bar specimen printing since it could reduce the residual stress and improve the overall build quality based on the literature report [43]. The laser scan angle at each layer can be retrieved directly from the LPBF machine during the hatching strategy setup. However, the laser scan might deviate from the specification during a real LPBF process due to possible flaws in galvanometer scanner control or errors in f-theta lens deflection. To obtain realistic hatching information, we measure the actual hatching angle by estimating it from the registered MP signature (temperature or area) maps as detailed in our previous works for this same batch of 16 fatigue bar samples [30]. It should be noted that the scan angle with a difference of 180° forms identical hatching pattern and strip overlap, which means a scan angle of 0° defines the same pattern as a scan angle of 180°. For this reason, the calibrated scan angle of the printing is obtained by subtracting 180° for scan angle larger than 180° (as shown in the bracket in Table 3). As shown in Table 3, similar and extremely high surface roughness values are present at Layer 67 and Layer 75 that have similar hatching angle of 141° and 137° (or 317°)**,** respectively. These two layers are exactly the same layers that have significantly more spatters per MP as observed in the previous section (Figure 15)**.** One possible reason for these two layers having the most spatters per MP and the highest surface roughness is that they undergo similar impacts of gas flow that will induce similar spattering, given their similar laser scan angle relative to the gas flow direction - 51° at Layer 67 and 47° at Layer 75. In this specific 16 fatigue specimen printing scenario, both the spatter count and surface roughness increase when the laser scan direction is around ~50 ° relative to the gas flow direction. At this hatching angle, more spatters are ejected, and irregular surfaces will be formed. Overall, spatter count is found to be partly attributed to the hatching angle, which essentially indicates the gas flow impact on spatter generation at the layer.

Table 3. Hatching angle for Layers 66, 67, 68, 69, 72, and 75 and the corresponding spatter count and surface roughness

| Layer # | Hatching Angle | Avg Spatter number per MP | Avg Sa ($\mu m$) |
|---|---|---|---|
| Layer 66 | 74° | 2.79 | 7.84 |
| Layer 67 | **141°** | **4.45** | **18.09** |
| Layer 68 | 208° (28°) | 3.01 | 10.27 |
| Layer 69 | 275° (95°) | 3.06 | 10.63 |
| Layer 72 | 116° | 2.61 | 14.77 |
| Layer 75 | **317°(137°)** | **4.23** | **19.60** |

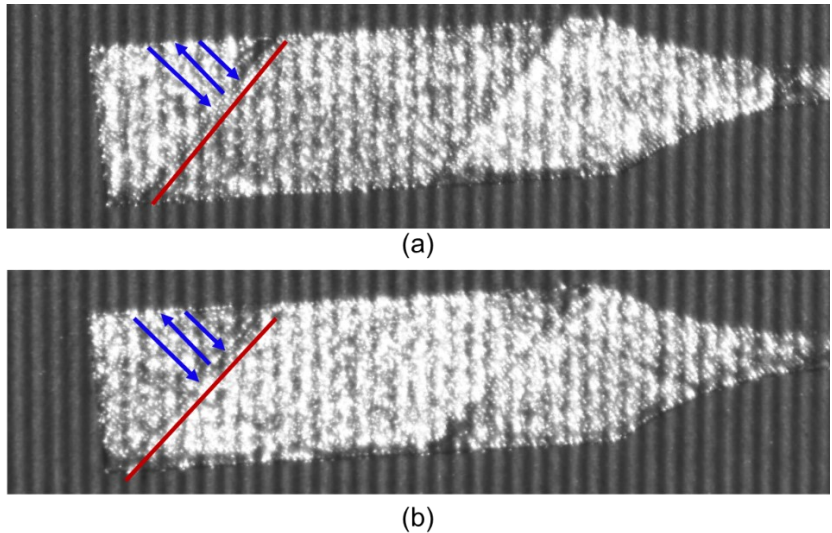

(a)

(b)

Figure 16. Fringe projection profilometry acquired images for Layer 67 (a) and Layer 75 (b). The hatch strip overlapping is annotated as red line and blue arrow is the laser scan pattern

To conclude, it is observed that in addition to laser power and speed (Section 3.3), scan angle or hatching angle also has a significant impact on spatter generation and in-process layer's surface quality. In other words, spatter signature could be an extensive barometer of dynamic changes in laser power, speed, and hatching angle. The advantageous benefit of using in-situ monitored spattering signatures instead of nominal LPBF process parameters as a predictor of layer surface quality is demonstrated in Section 3.5.

### 3.5 Correlation between spatter and in-process layer surface roughness

In Sections 3.2 – 3.4, we first find a qualitative correlation between the spatter count and in-process layer surface roughness; then we see that laser power, speed, and hatching angle can significantly affect the spatter count. One may argue that we may directly use priori information of nominal process setting to predict the spatters and thus the layer surface roughness without needing any in-situ sensing equipment and data analysis work specifically for spatter monitoring and correlation to surface properties. In this section, we quantitatively evaluate the influence of spattering on layer surface roughness as well as elucidate the importance of measuring spatter signatures in situ via regression modeling analysis using different combinations of inputs - average spatter count per MP and key process parameters including laser power (P), scan speed (V), HS, or VED, and hatching angle.

Specifically, seven SVM models using different combinations of inputs and a gaussian kernel function are trained and compared. The seven trained SVM models are applied to predict the surface roughness of each of the six sample layers (i.e., L66, L67, L68, L69, L72, L75) in the six FPP-monitored fatigue bars (Bars 2, 4, 6, 8, 10, and 12, as detailed in Section 3.1). Relative error is calculated between the predicted surface roughness and the actual surface roughness estimated using the FPP measurement results in Section 3.1. The relative errors of all the 252 model predictions (7 model prediction/layer × 6 bar × 6 layer/bar) accuracy are compared in Figure *17*, which shows that across all the layers and bars the models using spatter count as input generally and significantly outperform those with input of only process parameters. The best model, shown using a mark of green downward triangle, adopts VED and spatter count as the SVM inputs. Among all the predicted results, the surface roughness predicted for Layer 66 has the highest relative error. The reason could be attributed to imbalanced dataset as the model is constructed on sparse data with six layers data. Oversampling and down sampling from minority class can happen and contribute to the predictions error.

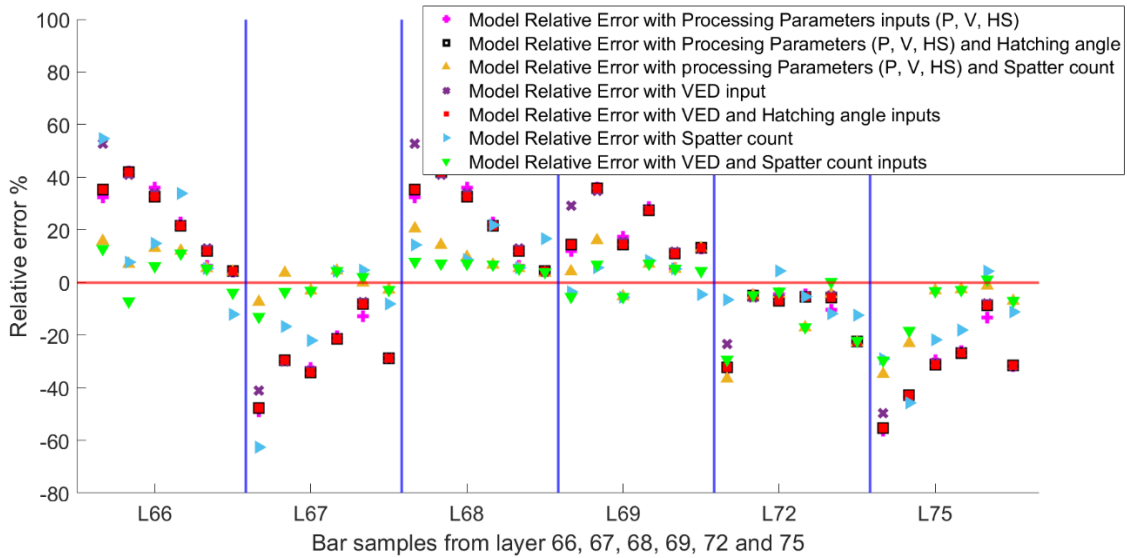

Figure 17. Relative error of the trained models using different combinations of features (process parameters and spatter signature) as predictors to predict the in-process layer surface roughness of six sample layers in six fatigue bars measured by the in-situ FPP. The division of the layers is annotated as the blue vertical line.

To further evaluate the performances of models with different inputs, three explicit metrics - Root Mean Squared Error (RMSE), Mean Absolute Error (MAE), and Mean Relative Error (MRE) are used. Table 4 shows the evaluation metrics of all the seven models. Based on the results, it is evident that using solely nominal processing parameters are not as accurate as using solely spatter count to predict layer surface roughness. Note that the two models of "(P, V, HS) + Hatching angle" and "VED + Hatching angle" have the same values due to their little difference (< 0.00002% relative error). However, comparing other pairs (marked with a matching color in Table 6) of models that use (P, V, HS) and VED plus the same input feature(s), we find that the models with VED would perform slightly better than those with (P, V, HS). Therefore, we recommend using VED as a LBPF process feature rather than (P, V, HS) to predict layer surface properties. This recommendation is also because VED (Equation (2)) incorporates the variable of layer thickness that can significantly affect spatter formation [42]. Moreover, models using hatching angle plus nominal process settings – (P, V, HS) or VED as the input perform better than those using only the process settings but are not as accurate as those using spatter count. We also attempt to purely use spatter count as an input without any process parameters and find that the model yield less error than those models using process parameters or VED with/without hatching angle. Further, the best model is found to be the one using spatter count along with VED, resulting in the lowest error in each of the metrics - RMSE, MAE, and MRE. All the results validate that spatter count is the most informative feature in this case due to its ability to reflect the influence of laser power, scan speed, and hatching angle. Our work also corroborates VED as an appropriate lumped, scaling factor for characterizing LPBF processes and incorporating the impacts of laser power, scan speed, hatching space, and layer thickness.

Table 4. SVM regression models' performance metrics for layer surface roughness predictions using different combinations of inputs. In-situ monitored signatures of Hatching angle and Average spatter count are boldfaced

| Model Input | RMSE ($\mu m$) | MAE ($\mu m$) | MRE (%) |
|---|---|---|---|
| Process parameters (P, V, HS) | 5.2 | 4.0 | 29.9 |
| Process parameters (P, V, HS) **Hatching angle** | 4.8 | 3.5 | 23.8 |
| Process parameters (P, V, HS) **Average spatter count** | 2.7 | 1.7 | 13.0 |
| VED | 5.1 | 4.0 | 29.8 |
| VED **Hatching angle** | 4.8 | 3.5 | 23.8 |
| **Average spatter count** | 4.2 | 2.7 | 18.7 |
| VED **Average spatter count** | **2.2** | **1.4** | **11.0** |

Besides, Wang et al [44], develops an analytic equation to predict the upper surface roughness of printed part using processing parameters and simulated MP depth information. Their model achieves the accuracy around 17.2% MRE. Our resulting model using VED along with the average spatter count per MP derived from in-situ monitoring data outperforms the literature analytic model with better accuracy (11.0% MRE) and computational efficiency (no need for compute-intensive simulation).

Based on all the quantitative comparisons above, we conclude that spatter monitoring and signatures registration can greatly help predict in-process layer surface properties more accurately. It is worth pointing out that our registered spatter signatures can aid in the prediction of both dynamic in-process layer surface roughness and as-built part properties.

## 4. Conclusion and recommendations

Spatters in LPBF processes tend to induce rough powder layers and printed layers, making LPBF printed components prone to porosities, cracks, and fractures. In this work, we conduct a systematic study from developing a framework of spatter monitoring and registration methods to quantify the effect of agglomerated powder and liquid droplet spatters on in-process layer surface roughness during LPBF. We find that the attained MP spatter feature profile can help predict the layer's surface roughness more accurately, in contrast to the traditional approaches that would only use nominal process setting or simulation without insights of real process dynamics. This is because the spatter information can reflect key process changes including the deviations in actual laser scan parameters (e.g., laser power, scan speed, hatching angle) and their effects. The results also corroborate the importance of spatter monitoring and the distinct influence of spattering on layer surface roughness. Our work paves a way for thoroughly elucidating the significant role of MP spattering in defect formation during LPBF and realizing online control and qualification of LPBF-AM processes.

Overall, the significant outcomes of this work can be summarized as follows.
- For spatter monitoring image processing and analysis, a deep learning-based image segmentation model using DeepLabV3+ plus Resnet is demonstrated to reduce the misclassification errors caused by camera lens flare and sensor noise and thus extract spatters from in-situ off-axis camera monitored images with good accuracy (99.5%).
- For spatter metrics extraction and registration, an unsupervised clustering method (DBSCAN) is employed to identify and count the continuous spatters. Ejection angle of each detected spatter is calculated. Due to the hardware limitation, only the liquid droplet spatters are characterized. Since liquid droplet spatters mainly occur at the rear side of MP, there is no significant variation of spatter ejection angles among different processing regimes. One can use a more capable camera setup to visualize other spatters whose ejection angles would differ with the process regime.
- A machine learning-based spatter features extraction and registration is developed using the methods as listed above. It can be extended to register more comprehensive metrics of all types of spatters given a better camera and improved hardware setup in the future.
- The registered spatter signatures reveal that the spattering phenomena especially the spatter count can change remarkably at different layers. Our results indicate that spatter formation is strong related to the LPBF processing

- parameters including not only laser power and speed but also hatching angle (essentially gas flow). With all the other conditions being the same, a linear correlation model between the spatter count and the laser power or the scan speed, respectively, can be established (Section 3.3). However, the spatter count could vary significantly across layers with the rotation of hatching pattern (Section 3.4), despite the layers being printed under the same nominal setting of laser power and scan speed. It is found that the layers with a scan angle of ~ 50° relative to gas flow direction incur more spatters per MP and higher surface roughness. As such, the developed spatter monitoring and signatures registration framework can provide quantitative insights to evaluate and compare different hatching strategies.
- The most accurate SVM-based regression model to predict in-process layer surface roughness is found to be the one that uses an input of both the spatter count derived from in-situ monitoring data analysis (Sections 2.2, 2.3, and 3.1) and the VED as features. It can efficiently predict the layer surface roughness with the least error (11.0% MRE) compared to the conventional approach of using nominal processing parameters as an input (29.8% MRE).
- All our experiment results show that only using nominal process setting (e.g., laser power, scan speed, or VED) and hatching angle is not sufficient to predict spatter features, much less layer surface roughness, necessitating the research on in-situ spatter monitoring and spatter-surface relationship modeling as demonstrated in this work.
- Incorporating the effects of laser power, scan speed, and scan direction relative to gas flow, the metric of spatter count is shown to be a potential LPBF performance indicator for predicting in-process layer surface properties more accurately than using nominal process parameters as demonstrated in this work (> 50% less error comparing the RMSE value of 2.2 μm vs ~5.0 μm as shown in Table 4). Provided enhanced monitoring systems (e.g., higher-speed camera that can capture the spatter redeposition location), more comprehensive spattering metrics can be obtained to serve as a set of inclusive, powerful process signatures for effectively predicting the properties of both in-process layer surface and final parts, including defects on exterior surfaces and inside the body.

Moving forward, our developed in-situ MP spatter monitoring system and registration framework can be further developed to provide spatial profiles of key spatter signatures for all monitored layers in LPBF, and our spatter-layer-surface-roughness correlation method can be used to aid process control or in-process defects correction for enhanced part properties. In the future, one can extend our spatter registration framework to include more types of spatters with advanced hardware setup. It is worth pointing out that our monitoring system does not track the full trajectory of spatter in this work. Besides, our method does not count the spatters overlapped with MP since the clustering method can only characterize disconnected parts. This limitation needs to be addressed in our future work. Nevertheless, we demonstrate the feasibility of effectively predicting the layer surface roughness – a global metric of surface property – by using only the spatter count from a subset of MP without knowing the exact landing location of spatters. Meanwhile, the in-situ FPP system can also be upgraded to address the phase error induced harmonic fluctuations in height profile. Especially, considering the unique LPBF processing conditions our group has recently improved the accuracy of existing FPP methods by incorporating build-location-dependent material reflectivity and sensor nonlinearity into the FPP measurement model [27]. The advancement of our LPBF-specific FPP technology will provide a more accurate approach to characterize the in-process layer surface roughness and analyze its correlation with the spattering phenomenon in LPBF based AM. Moving forward, we will improve and implement the monitoring setup to capture the full spatter trajectory and to measure the layer surface with better accuracy and higher resolution so that we can elucidate the build location effect on spatter generation as well as the localized effect of spattering on the pixel-resolved surface properties and further predict the final part's internal defects such as porosity and micro-crack. Furthermore, LPBF process optimization and control can be conducted by using the dynamic feedback of spatter signatures and layer surface roughness to improve the final printed part properties.

## Acknowledgements

Authors acknowledge the partial support from the Department of Energy University Turbine Systems Research grant (Award #: FE0031774), as well as the support from the National Science Foundation (NSF) sponsored industry/university cooperative research center (IUCRC) – the Center for Materials Data Science for Reliability and Degradation (MDS-Rely) (Award #: 2052662). Any opinions, findings, and conclusions or recommendations expressed in this publication are those of the authors and do not necessarily reflect the views of the NSF. This work used the Extreme Science and Engineering Discovery Environment (XSEDE) resource - Pittsburgh Supercomputing Center Bridges-2 GPU and Storage through allocation MCH210015. Authors would also like to thank the facilities support from ANSYS Additive Manufacturing Research Laboratory (AMRL) at the University of Pittsburgh and Brandon Blasko for his help with experiment.

## CRediT Author Statement

**Haolin Zhang:** Methodology, Software, Investigation (Experiments and Data acquisition), Data Analysis, Writing - Original draft preparation & Editing, and Visualization.

**Chaitanya Krishna Prasad Vallabh:** Methodology, Software, Investigation (Experiments and Data acquisition), Data Analysis, and Writing- Draft, Review & Editing.

**Xiayun Zhao:** Conceptualization, Methodology, Resources, Writing – Draft, Review & Editing, Supervision, Funding acquisition, and Project administration.

## Appendix

### A1. Dilated convolution for semantic segmentation neural networks

The primary objective of using the dilated convolution is to expand and extract features from different field of view as the convolutional filter is dilated with certain stride for the segmentation neural network. Dilated convolution is the convolution filter which is convenient in controlling the resolution of features extracted and the filter's field of view. This filter is powerful in augmenting features from broader view. Specifically for the image data, the output feature map $y$ from convolution operation is computed using input feature map $x$ and convolution filter $w$.

$$y_i = \sum_{k} x_{i+r \cdot k} w_k \qquad \text{A-(1)}$$

In Eq. A-A-(1), $r$ is the dilated stride or rate, and $k$ is the convolution size. When the dilated stride is 1, the convolution becomes the standard depth-wise convolution kernel filter. The primary objective of using the dilated convolution is to expand and extract features from different field of views for the segmentation neural network.

### A2. Experiment data of spatter count per melt pool under different settings of laser power and scan speed

Table A-1 and Table A-2 present the average spatter count per MP for Layer 66 at varying laser and scan speed, respectively, while keeping all the other processing parameters the same. It can be observed that the spatter formation is linearly related to both laser power and scan speed with all the other conditions remaining the same.

Table A-1. Average Spatter Count at varying laser power at L66

| Laser Power | Average Spatter count in a print layer |
|---|---|
| 200 W | 1.4 |
| 250 W | 2.3 |
| 300 W | 3.6 |
| 350 W | 4.1 |

Table A-2. Average Spatter Count at varying scan speed at L66

| Laser Scan Speed | Average Spatter count in a print layer |
|---|---|
| 0.5 m/s | 3.0 |
| 0.75 m/s | 2.3 |
| 1.0 m/s | 1.0 |